\documentclass[aps,prl,twocolumn,showpacs,superscriptaddress,groupedaddress,nofootinbib]{revtex4-1}  
\usepackage{graphicx}  
\usepackage{dcolumn}   
\usepackage{bm}        
\usepackage{amssymb}   
\usepackage{amsmath}
\usepackage{xspace}
\usepackage{epstopdf}
\usepackage{wrapfig}
\usepackage{color}
\usepackage{bbm}
\usepackage{float}

\usepackage{verbatim}

\usepackage{capt-of}

\usepackage{ulem} 

\newcommand{\beq}{\begin{equation}}
\newcommand{\eeq}{\end{equation}}
\newcommand{\bea}{\begin{align}}
\newcommand{\eea}{\end{align}}

\newcommand{\be}{\begin{equation}}
\newcommand{\ee}{\end{equation}}

\newcommand{\s}{$S$}

\newcommand{\ODMOb}{$\Omega_{DM}/\Omega_b$}
\newcommand{\yb}{$y_b$}

\begin{document}
\title{
A precision test of the nature of Dark Matter\\ and a probe of the QCD phase transition}
\author{Glennys R. Farrar}
\date{March 9, 2018}   
\affiliation{Center for Cosmology and Particle Physics,
Department of Physics,
New York University, NY, NY 10003, USA}

\date{Sept. 9, 1018, expanding on original version of May 9, 2018}
\begin{abstract}  
If dark matter (DM) contains equal numbers of $u,d,s$ quarks, the ratio of DM and ordinary matter densities after hadronization follows from the Boltzmann distribution in the Quark Gluon Plasma.  For sexaquark $uuddss$ DM in the 1860-1880 MeV mass range (assuring sexaquark and nuclear stability) and quark masses and transition temperature from lattice QCD, the observed $\Omega_{DM}/\Omega_b = 5.3$ is in the predicted range, with $\lesssim 15$\% uncertainty.  The prediction is insensitive to the current form of DM, which could be sexaquarks, strange quark matter nuggets, primordial black holes from their collapse, or a mix.  
\end{abstract}
\maketitle 

\section{Introduction}
\label{sec:intro}
Ordinary matter (OM) is composed almost entirely of nucleons, by mass, and without a baryon chemical potential there would be a negligible residual nucleon density after matter-anti-matter annihilation \cite{KolbTurner}.  The nucleon-to-photon ratio\footnote{Following conventional notation, subscript $b$ denotes nucleons.} $\eta \equiv n_b/n_\gamma = 6.58\pm 0.02 \times 10^{-10}$ is measured via cosmic microwave background (CMB) tracers of recombination and through primordial ``Big Bang" nucleosynthesis (BBN) \cite{pdgCosmo17}, with their consistency being a triumph for the Standard Cosmological Model.  $\eta \ne 0$  reflects the baryon asymmetry of the Universe (BAU); accounting for its value is a profound challenge to theory.   
In most DM models the DM relic density is  unrelated to the baryon asymmetry and results from some entirely different process such as annihilation of DM particles (WIMP and other thermal-relic scenarios) or non-thermal processes (axion-DM).  
In such models the similarity of mass densities of DM and ordinary matter, $\Omega_{DM}/\Omega_b = 5.3 \pm 0.1$\cite{PlanckCosmo15}, is just an accident \cite{SM1}. 

Here I show that the observed value of $\Omega_{DM}/\Omega_b $ follows from the Boltzmann distribution in the Quark-Gluon Plasma  (QGP) with minimal additional assumptions, as long as DM is composed of $u,d,s$ quarks.  This follows largely independently of the present form of DM which could be stable sexaquarks\footnote{So named because Greek prefixes are used for hadrons with an extra $q\bar{q}$ pair such as tetra- and pentaquarks, and sexa is the Latinate cardinal prefix for six;  see the table at https://en.wikipedia.org/wiki/Numeral\_prefix and \cite{fSexaquark17}.} \cite{fSexaquark17}, quark nuggets  \cite{wittenQuarkNuggets84}, primordial black holes (PBH) formed from either of these, or a combination of them.

After briefly reviewing the sexaquark and quark nugget models and the QCD phase transition, the main result is derived: an expression for \ODMOb\ which depends on the quark masses and the temperature at the end of the QGP phase, the Dark Matter mass per unit baryon number, \yb, and the efficiency with which strange quarks are entrained into dark matter.
These factors are well enough determined from statistical physics and knowledge of the hadrons and lattice QCD, that \ODMOb\ can be predicted to within a factor-few uncertainty.  The prediction agrees well with observation.   The total baryon number asymmetry, including that contained in DM, is $\eta_{\rm tot} = \eta (1 +  \Omega_{DM}/( y_b \Omega_b )) \approx 4.1 \times 10^{-9}$.  How this asymmetry arises remains to be explained, but presents no greater challenge than in other DM models.

\section{Models with $uds$ Dark Matter}
\label{udsDM}
As will be explained below, the value of \ODMOb\ which results when DM is made of $uds$ quarks is only weakly dependent on the form the DM takes, whether particulate DM or massive compact objects.  Here we briefly review some of the options.

The stable sexaquark hypothesis \cite{fSexaquark17} postulates that the Q=0, B=+2, $uuddss$ flavor-singlet scalar bound state (denoted \s) is  stable.  The \s\ is absolutely stable if $m_S  \le 2 \, (m_p + m_e) = 1877.6$ MeV.  A somewhat higher mass may also possible, because up to $m_S = m_p+m_e + m_\Lambda = 2054.5 $ MeV the \s\ decays through a doubly-weak interaction and, if sufficiently compact, its lifetime may be longer than the age of the Universe \cite{fzNuc03}.  Both cases are called ``stable" below for conciseness.   The $S$ cannot be too light, or nuclei would decay.  The most constraining process is $n p \rightarrow S e^+ \nu_e$ because $n n \rightarrow S \gamma$ is highly suppressed due to the \s\ being uniformly neutral and having no magnetic moment \cite{fSexaquark17}.  Nuclei are of course stable if $m_S > m_p + m_n - m_e  - 2 BE$, where $2 BE$ is the binding energy of the $n+p$.  Thus $m_S \gtrsim 1861$ MeV does not threaten nuclear stability\footnote{Primordial D/H is within errors for $\tau_d \approx \tau_n /[ G_F \,m_p^2 \,\,{\rm sin} \theta_C \times ({\rm wavefunction \,overlap})]^2 \gtrsim 5 \times 10^{14}$ yr.  This is easily satisfied because hypernuclear experiments imply the $({\rm wavefunction \,overlap})^2 \lesssim 10^{-8}$ \cite{fzNuc03}.  Sudbury Neutrino Observatory threshold was 5 MeV, restricting reach to $m_S \lesssim 1870$ MeV;  refined sensitivities and rates for nuclear processes in \cite{fw18inprep}.}.   

Initial searches for a $uuddss$ bound state were stimulated by Jaffe's MIT bag model estimate of 2150 MeV \cite{jaffe:H}.  With a mass below $2 m_\Lambda = 2230$ MeV, the state would be strong-interaction-stable and have a lifetime $\gtrsim 10^{-10}$s;  Jaffe called the state  ``H-dibaryon".  
Although its mass was uncertain, the H-dibaryon was almost universally assumed to decay weakly as a result of thinking $m_H > m_p + m_\Lambda (2054$ MeV).  Additionally, the H was envisaged as a loosely-bound di-$\Lambda$, readily formed in hypernuclei, e.g., \cite{Baym+JaffeCygX3_85}.   Dozens of experiments were performed attempting to find an H-dibaryon, and the ensemble of null-results has widely been taken to exclude it.   

While the original H-dibaryon is likely excluded, a careful re-examination of the experimental situation \cite{fSexaquark17} shows that no experiment to date would have detected a compact, stable \s.  Experiments either required $m_H > 2$ GeV, or searched for a signal in the invariant mass of decay products such as $\Lambda p \pi^- $, or implicitly assumed a dibaryon spatial configuration comparable to a deuteron or nucleon so its interactions and production was expected to be comparable to ordinary hadrons; see \cite{fSexaquark17} for further discussion.

The stable sexaquark hypothesis is tenable due to the unique symmetry of the $uuddss$ ground state.  Models designed to fit known hadrons cannot be trusted to reliably describe it because Fermi statistics prevents mesons and baryons from enjoying the triply-singlet configuration (in color, flavor, spin) accessible to $uuddss$.   Hyperfine attraction is strongest in singlet configurations, c.f., the Most-Attractive-Channel hypothesis \cite{MAC}, so binding is maximal in the sexaquark channel.  Furthermore, the \s\ should be much more compact and weakly coupled than normal hadrons due to being a flavor singlet and thus not coupling to pions.  Baryons ($r_N = 0.9$ fm) are much larger than their Compton wavelength ($\lambda_N = 0.2$ fm), which can be attributed to baryons coupling to pions ($\lambda_\pi = 1.4$ fm).   Estimating $r_S = \lambda_S + (0-0.5) \lambda_{M1}$ by analogy with baryons, where $M1$ is the lightest well-coupled flavor singlet meson (the $f_0$ or $\omega-\phi$ with $m_{M1} \sim 500-1000$ MeV), gives $r_S = 0.1-0.3$ fm.  A compact, stable \s\ is thus both self-consistent and  phenomenologically allowed, because the disparate size of \s\ and baryons means amplitudes involving overlap of \s\ and two baryons are very suppressed;  see \cite{fSexaquark17} for more details. 

The failure to find an H-dibaryon, combined with the well-founded expectation that some bound state -- either stable or decaying -- should exist, along with the absence of searches which could have detected a stable \s, is strong indirect evidence for an as-yet-unobserved stable \s\ with mass $\lesssim 2$ GeV.   New experimental strategies suited to finding such a particle, which is surprisingly elusive, were outlined in \cite{fSexaquark17}; experimental searches are underway.  

If a stable \s\ exists, it could be the DM.  Lacking coupling via pions, $\sigma_{S N}$ is lower or much lower than for ordinary hadron scattering, e.g., $\mathcal{O}(10^{-30} {\rm cm}^2)$, implying the \s's astrophysical and cosmological impacts are probably negligible \cite{fSDM17}.  Direct detection constraints on hadronically interacting DM (HIDM) are limited to $m_{\rm DM} \gtrsim 2$ GeV \cite{mfVelDep18,fx18inprep}.   The best limits on DM in the mass range $\approx 0.6-6 $ GeV are the indirect limits of \cite{nfm18}, from HST orbital decay and evaporation of liquid cryogens, and thermal conductivity of the Earth.  These limits remain compatible with 100\% of the DM consisting of \s\ \cite{fx18inprep}.  Limits from the cosmological power spectrum extend to lower mass but are compatible with expectations for \s DM:   for 1 GeV DM \cite{gluscevicBoddy17,xuDvorkin+18}, $\sigma_{{\rm DM}p} \lesssim 5 \times 10^{-26} {\rm cm}^2$.

The quark nugget scenario was inspired by the possibility that a macroscopic state of strange quark matter could be lower in energy than a non-strange state of equivalent baryon number \cite{wittenQuarkNuggets84,farhiJaffeStrangeMatter}.  Witten showed that if the QCD phase transition is first order, quark nuggets might form \cite{wittenQuarkNuggets84}.   We now know from lattice QCD \cite{hotQCD14} that the transition is not a first order phase transition.  Nonetheless some nuggets may form --  perhaps via condensates of sexaquarks seeded by primordial density inhomogeneities -- although the large background of $q \bar{q}$ pairs would appear to make this difficult.  If some quark nuggets do form, all or a portion of them might collapse to primordial black holes (PBH) of the same total mass.   PBH have been proposed as a component of dark matter, to help explain the very high mass supermassive black holes observed already at high redshift, and as a possible explanation for the large masses of LIGO's black hole merger events \cite{LIGOdm16}.  

As we shall see, the value of \ODMOb\ derived below is insensitive to what form the DM has in the Universe today, as long as it was assembled from equal numbers of $u,d,s$ quarks.  In the sexaquark model the numbers of $u,d,s$ are exactly equal, while in the quark nugget model the numbers are equal except in a thin surface layer.   The mass-per-unit baryon number of the DM, $y_b \, m_p$, is very close to $m_p$ for sexaquarks, where the mass range $1860 < m_S <1900 $ MeV implies $y_b = m_S/(2 m_p) \approx 0.99 -1.01$, and $y_b$ would be similar for quark nuggets if those exist \cite{farhiJaffeStrangeMatter}.   If agglomerations of sexaquarks or quark nuggets further evolve to become PBHs, the total mass in PBH would be essentially the same as that of the $u,d,s$ material from which it formed, since the sexaquarks and quark nuggets have negligible coupling to photons and the collapse process can be expected to be nearly spherically symmetric and thus not emit gravitational radiation.  

\section{QCD phase transition}
\label{QGPtran}
At high temperature, the QCD sector consists of a plasma of massless gluons, nearly massless $u, \bar{u},d,\bar{d}$ quarks and intermediate mass $s,\bar{s}$ quarks.   At low temperature, the QGP is replaced by the chiral-symmetry-broken, color-confined phase in which baryons are heavy and pseudoscalar mesons are light.   Lattice QCD calculations show that the transition between the QGP and the low temperature hadronic phase is a cross-over centered on 155 MeV \cite{hotQCD14} rather than a true phase transition.  As the temperature drops from 170 MeV to 140 MeV, the quark and gluon condensates responsible for hadron masses and color confinement increase; at the same time it becomes more favorable energetically for $q \bar{q}$'s and $qqq$ to combine into color singlet mesons and baryons.   Typical intra-$q, \bar{q},g$ separations (plotted in Fig. \ref{sep} in the Supplemental Materials) are $\mathcal{O}(1$ fm) for $T \approx 150$ MeV.  The age of the Universe in this epoch is 
$
t_{\rm Univ} = 7.3 \times10^{-5} (100 {\rm\, MeV}/T)^2 \, {\rm sec}, 
$
whereas the timescale for hadronic interactions is $\mathcal{O}(10^{-23}$s), so chemical as well as thermal equilibrium is maintained.

The equilibrium number density of each fermion species as a function of temperature is given by 
\be
\label{nq}
n(m,T) = \frac{g}{2 \pi^2} \int_m^\infty \frac{E \sqrt{E^2 - m^2}}{e^{(E \mp \mu)/T} + 1}\, dE, 
\ee
where $g$ is the number of degrees of freedom (2-spin$\times$3-colors $\rightarrow g= 6$ per $q$ and $\bar{q}$ flavor) and $\mu$ is the chemical potential.   A plot of the equilibrium abundance of various species as a function of temperature, is provided for reference in Fig. \ref{ratio} in the Supplemental Materials.

The quark masses are accurately known from the hadron spectrum in lattice QCD \cite{bazazov+QuarkMasses18}: $m_u = 2.118(38)$ MeV, $m_d = 4.690(54)$ MeV and $m_s = 92.52(69)$ MeV.  In the QGP, the relative abundances of photons, gluons, and light quarks $u, \bar{u},d,\bar{d}$ are in the ratios 1:8:$\frac{9}{4}$, and $s$ quarks have a slightly lower abundance.  These flavor ratios apply both to the thermal $q \bar{q}$ quarks and the ``baryon excess" quarks.   The BAU amounts to a roughly part-per-billion difference between the $q$ and $\bar{q}$ abundance for each light flavor; to excellent approximation the chemical potential can be ignored above 100 MeV.    Below the hadronization transition, the most abundant particles besides photons and leptons are pions (c.f., SM Fig. \ref{ratio}).  Weak interactions maintain flavor chemical equilibrium, and hadronic and EM reactions like $\pi^+ \pi^- \leftrightarrow \gamma \gamma$ keep hadron abundances in thermal equilibrium well into the low temperature phase.  

\begin{figure}[H]
\centering
\vspace{-0.1in}
	\includegraphics[width=0.9\linewidth]{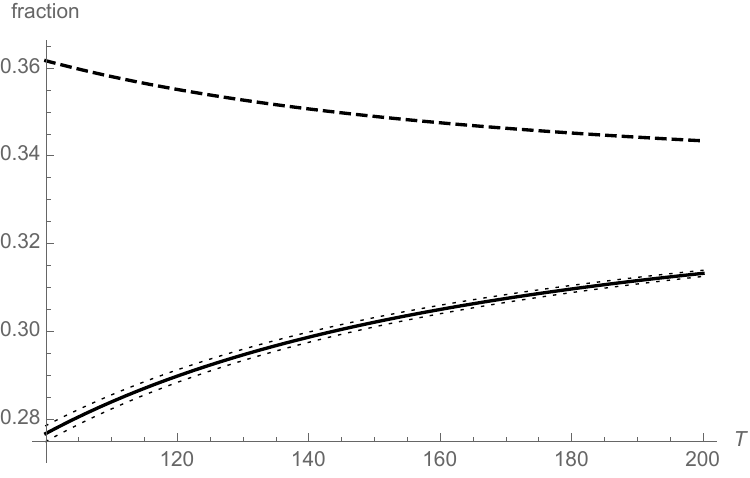}
	\vspace{-0.1in}
		\caption{Fractions of quarks in thermal equilibrium that are $s$ (solid) and $u,d$ (dashed), as a function of $T$ in MeV.  Dotted lines show the effect of a $\pm 3 \sigma$ shift around the central value of $m_s$; the corresponding smaller shifts for $u,d$ are not shown.}
	\label{fsfqvsT}
\end{figure}

\section{Dark-to-Ordinary Density ratio: $\frac{\Omega_{DM}}{\Omega_b}$}
\label{DMtob}
Simply due to their higher mass, the equilibrium fraction of strange quarks and antiquarks, $f_s \equiv (n_s + n_{\bar{s}})/\sum_{i=1}^3(n_i + n_{\bar{i}})$ is lower than that of up and down quarks and antiquarks, as shown in Fig. \ref{fsfqvsT}.  In the DM scenarios under consideration, the DM contains equal numbers of $u,d,s$ quarks and the maximum DM to OM ratio occurs when every $s$ is entrained in DM and the left-over $u,d$ quarks make baryons, leading to $3 f_s/(1 - 3 f_s)$ DM particles per unit baryon number.   As the temperature drops from 170 to 140 MeV, $3 f_s$ changes from 0.964 to 0.948 leading to \ODMOb$_{\rm max}$ = (18-27)$y_b$.  
However not every $s$ is entrained in DM so the actual ratio of dark matter and ordinary mass densities is
\be
\label{ODMoverOb}
\frac{\Omega_{DM}}{\Omega_b} = \frac{y_b \,  \kappa_s \, 3 f_s}{1 - \kappa_s\, 3  f_s}~,
\ee
where $\kappa_s $ is the efficiency with which $s$ quarks are trapped in DM at the end of the hadronization transition.  

We can estimate $\kappa_s $ as follows.  First consider production of \s's.  Even at the level of 1-gluon exchange, which provides a good qualitative accounting of most hadron mass splittings \cite{DGG75}, there is a strong hyperfine attraction between $uuddss$ quarks in the sexaquark (color-, flavor- and spin-singlet) configuration \cite{MAC,jaffe:H}.  This perturbative attraction is present independently of whether the quarks are in an isolated, zero-temperature \s\ particle, quark nuggets, or are in the QGP.   Thus when the strongly attractive sexaquark configuration of quarks occurs by chance in some spatial region of the QGP, it will be energetically favored and linger in that state.  Quarks in configurations which are not energetically favored will continue their random rearranging.  

Because the chemical potential is negligible, statistical physics tells us that the relative probability of finding two $s$ quarks in an \s-like state compared to finding them in a state consisting of two separate (hyperon-like) 3-quark states, is exp$(\Delta E)/T$ where $\Delta E$ is the energy splitting of the two configurations.   When hadronization occurs, the \s-like color singlet states become \s's and other color singlets become mesons, baryons and anti-baryons;  configurations which are not color singlets continue rearranging and form new color-singlet combinations which then become hadrons.
Hyperons and anti-hyperons present after hadronization maintain their appropriate thermal equilibrium abundances by decaying or scattering into nucleons and anti-nucleons, or annihilating. 

We can estimate $\Delta E$ and hence $\kappa_s$ using physical masses of nucleons, hyperons and the hypothesized mass of the \s; this approximation gives
\be
\label{kappa}
\kappa_s(m_S,T) = \frac{1}{1 + \left( r_{\Lambda,\Lambda} +  r_{\Lambda,\Sigma} + 2  r_{\Sigma,\Sigma} + 2  r_{N,\Xi}\right)} ~,
\ee   
where $r_{1,2} \equiv {\rm exp}[-(m_1 + m_2 - m_S)/T] $ and the coefficients of the different terms are the number of combinations of the given baryon states containing $uuddss$.   The leading uncertainty due to confinement and chiral-symmetry breaking cancels, to the extent that the presence or absence of the quark and gluon condensates shifts the masses of the \s\ and octet baryons together.  

\begin{figure}[tbp]\centering
	\includegraphics[trim={0 0.05in 0 0},clip,width=\linewidth]{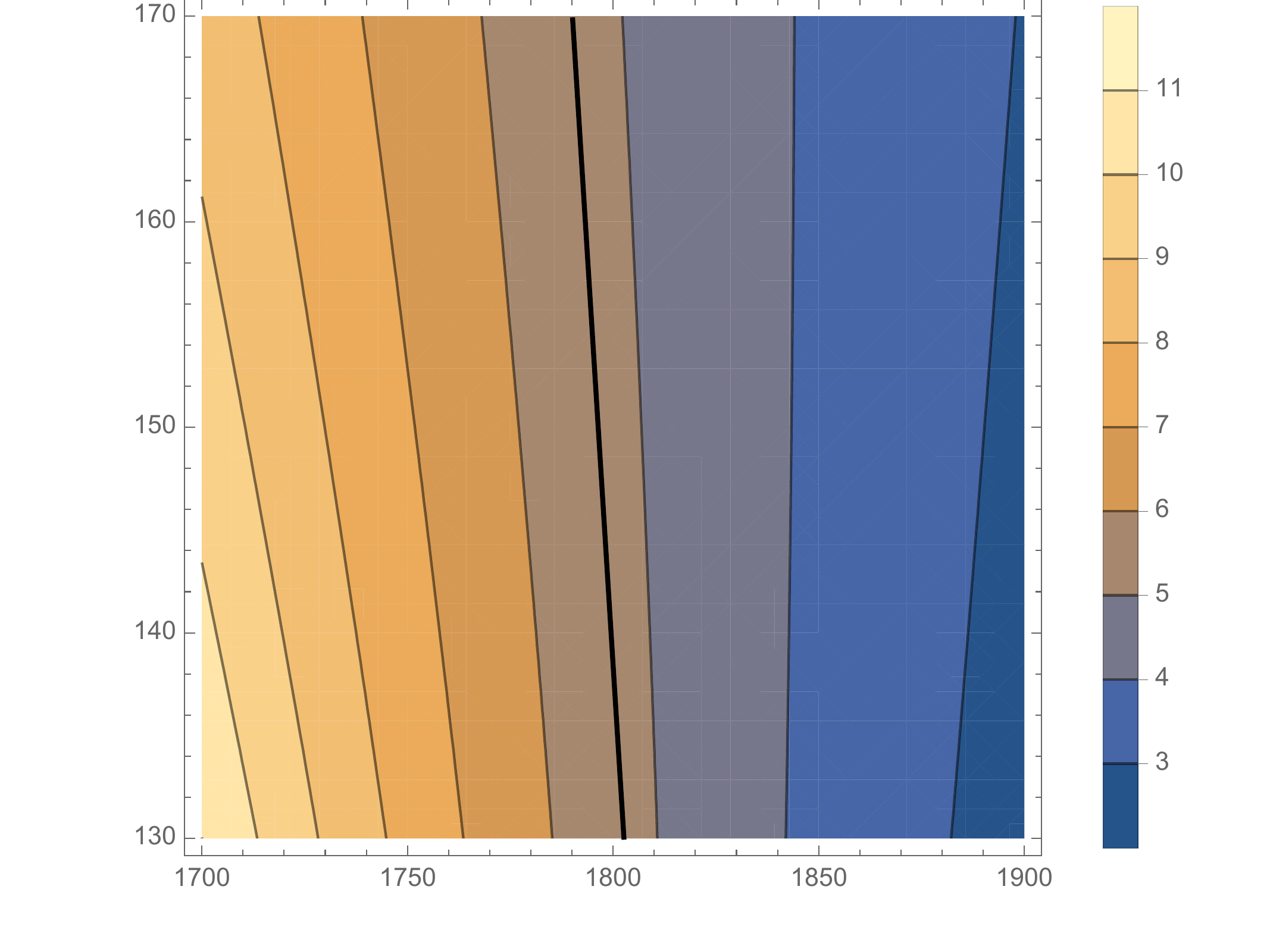}
		\caption{$\Omega_{DM}/\Omega_b$ versus $m_S$ in MeV (vertical axis) and the effective transition temperature in MeV (horizontal axis).  The measured value $5.3 \pm 0.1$ is indicated by the black line.  (Note there was a plotting bug in this figure in the original version of this preprint; this is the corrected plot.) }
	\vspace{-0.1in}
		\label{finalplot}
	\vspace{-0.1in}
\end{figure}

Idealizing the production of DM as occurring at a single effective temperature somewhere in the 140-170 MeV range, and using Eq. (\ref{kappa}) to calculate $\kappa_s(m_S)$ taking $y_b = m_S/(2 m_p)$, leads to the values of  \ODMOb\ shown in Fig. \ref{finalplot}.  For the entire plane the predictions are within a factor-2 of the measured ratio $\Omega_{DM}/\Omega_b = 5.3 \pm 0.1$.  Indeed, the agreement is better than 15\% for sexaquark mass in the ``safe" 1860-1880 MeV range, for the arguably most relevant range for $T_{\rm eff}: \, 140-155$ MeV, the final stage of hadronization.  

The mild dependence of \ODMOb\ across Fig. \ref{finalplot} follows from the fact that $f_s$ and $\kappa_s$ have the opposite behavior as $T$ changes so their product varies relatively little.   The biggest source of uncertainty in predicting \ODMOb\ is thus $m_S$, and the approximation of using $T=0$ values of the masses to estimate $\kappa_s$ via Eq. (\ref{kappa}).  If a sexaquark would be discovered so $m_S$ is fixed, the 2\% precision with which \ODMOb\ is known will strongly constrain how the QCD condensates and energy difference of sexaquark-like and hyperon-like states evolve with temperature.  If DM is made of $uds$ quarks but not sexaquarks, then the $m_S$ in Fig. \ref{finalplot} is to be interpreted as an effective $uuddss$ energy during DM formation, appearing in Eq. (\ref{kappa}) via the expression for $r_{1,2}$.

Implicit in the above discussion, is that the value of \ODMOb\ established in the hadronization transition persists to the recombination epoch where it is measured \cite{PlanckCosmo15}.   If DM is in the form of quark nuggets or PBH, its persistence is assured as long as the nuggets or PBHs are large enough not to evaporate.  (Late-time accretion of OM onto PBH, even if significant, would not affect the value of \ODMOb\ at recombination.)  For sexaquark DM, non-destruction requires the cross section for reactions such as $\pi S \rightarrow \Sigma \Lambda$ and $K S \rightarrow p \Lambda$ to be small, consistent with the wavefunction overlap between an \s\ and two baryons being suppressed due to the expected small size of the sexaquark \cite{fzNuc03, fSexaquark17};  see Supplementary Materials for more details .

\section{Conclusions}
Very simple statistical physics, along with externally-determined parameters such as quark masses and the temperature range of the QGP-hadron transition, have been used to calculate the ratio of Dark Matter to ordinary matter in models where DM is comprised of equal numbers of $u,d,s$ quarks.  The prediction is relatively insensitive to details of DM properties.  When applied to the sexaquark dark matter model, the prediction agrees with observation to within its $\approx15$\% uncertainty for $m_S$ in the range 1860-1880 MeV, where both sexaquark and nuclei are absolutely stable.  Subsequent breakup of sexaquarks into baryons must be suppressed, but this is self-consistent in the framework of the stable sexaquark conjecture due to its deep binding and compact physical size.

Conditions in the QGP phase transition may lead to multiple coexisting states of $uds$ dark matter, potentially consisting of a combination of sexaquarks, quark nuggets and primordial black holes \footnote{The microscopic details of the QCD phase transition and their interplay with primordial temperature/density fluctuations from inflation have to be explored in greater depth, to determine whether these inhomogeneities can produce regions in which the development of the chiral condensate lags sufficiently relative to that in neighboring regions, to significantly concentrate baryon number, along the lines considered in connection with a 1st order transition \cite{wittenQuarkNuggets84}.  An intermediate phase of sexaquarks or sexaquark-like states in the late-stage QGP as discussed in connection with Eq. (\ref{kappa}) might facilitate this process. }.   A similar if less precise prediction for \ODMOb\ as seen in Fig. \ref{finalplot} should apply to both quark nuggets and primordial black holes from their collapse, if either of those form, since they share the essential features:  {\it i)} DM which consists of nearly equal numbers of $u,d,s$ quarks and {\it ii)} relative suppression of hyperon-like states, due to the strong QCD attraction of $uuddss$ quarks in the spin-0 channel.      

The parameter-free nature and simplicity of the analysis presented here make the congruence of prediction and observation a possible smoking gun that DM consists of $u,d,s$ quarks in nearly equal abundance.  In that case, \ODMOb\ will provide a valuable window onto the QGP-hadron transition, much as the abundances of primordial nuclei probe conditions during nucleosynthesis at 1000 times lower temperature.  

The author is indebted to the Simons Foundation for a Senior Fellowship in Theoretical Physics, during which the foundations for the present work were laid, and to Y. Ali-Ha\"imoud, N. Christ, A. Grosberg, C. McKee, J. Ruderman and B.~J. Schaefer for helpful comments, information and discussions. 



\clearpage 
\newpage 

\section{Supplemental Materials}
\vspace{-0.1in}

\subsection{Relationship to the BAU}
In the present work, as in most efforts to explain the DM to OM ratio, a baryon asymmetry of the Universe (BAU) is taken as an input.  One exception to this is the proposal of Farrar and Zaharijas (2006), SM\cite{fz06} (FZ06 below), which seeks to explain the BAU by eliminating it.  In the scenario of FZ06 there is no excess of net baryon number, but instead anti-baryon number is preferentially sequestered in the dark sector.  A particularly satisfying realization of the scenario would be for the DM to be mostly anti-sexaquark (aka H-dibaryonic) because then as shown by FZ06 one could explain \ODMOb\ $\approx 5$ as well as the baryon asymmetry.  But FZ06 showed that any scenario explaining both the BAU and \ODMOb\ $\approx 5$ is excluded by limits on the internal heating of Uranus, unless the DM-nucleon annihilation cross section highly suppressed relative to the level expected from hadronically interacting DM.  Thus FZ06 provided a second realization of the anti-baryon sequestration mechanism which did not explain \ODMOb.  To summarize, the heating argument of FZ06 eliminates the possibility of using the anti-baryon sequestration mechanism of FZ06 to explain the BAU with anti-sexaquark DM, but does not impact the viability of sexaquark DM given a primordial BAU, which is the perspective of the present work. 

\subsection{Plots to aid intuition about the QGP transition}

\begin{figure}[H]
\centering
	\vspace{-0.4in}
	\includegraphics[width=0.9\linewidth]{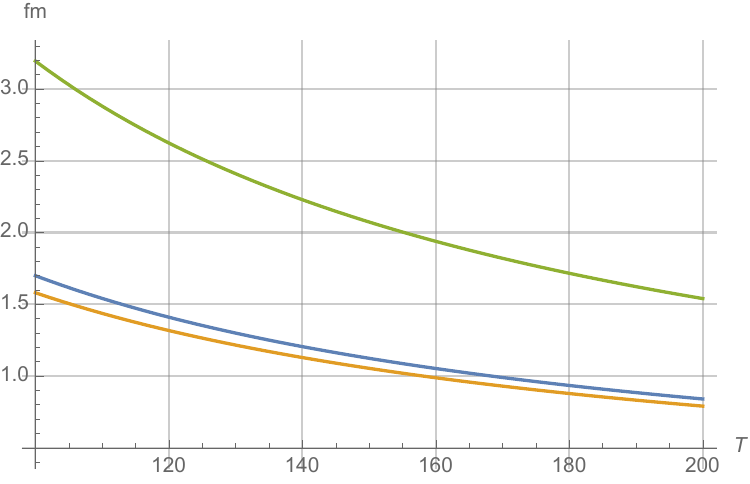}
		\caption{The mean separation between light quarks (blue), gluons (orange) and strange quarks (green) versus temperature in MeV; the QCD phase transition occurs over the temperature range 140-170 MeV.}
	\label{sep}
\end{figure}

\begin{figure}[H]
\centering
	\includegraphics[width=0.95\linewidth]{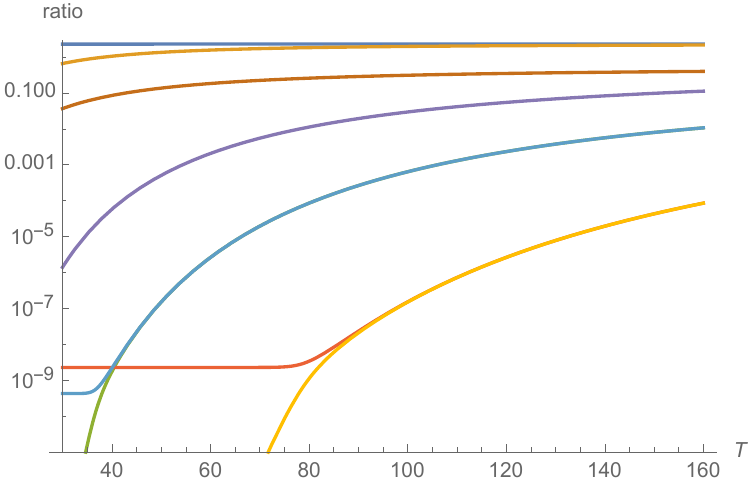}
		\caption{Number density ratio of various particles to photons, as a function of temperature in MeV, assuming the particle exists and is in thermal equilibrium at the given temperature, and that baryon number does not transfer between the \s\ and ordinary baryons as discussed below.   Top to bottom: $\frac{n_u}{n_\gamma}
		$ 
		(blue), $\frac{n_s}{n_\gamma}
		$
		(ochre), $\frac{n_{\pi^+}}{n_\gamma}$ (brown), $\frac{n_{K^+}}{n_\gamma}$ (purple), $\frac{n_{p,\bar{p}}}{n_\gamma}$ (turquoise/green) and $\frac{n_{S,\bar{S}}}{n_\gamma}$ (red/yellow).
		}
	\label{ratio}
\end{figure} 

\begin{figure}[H]
\centering
\vspace{-0.08in}
	\includegraphics[width=0.85\linewidth]{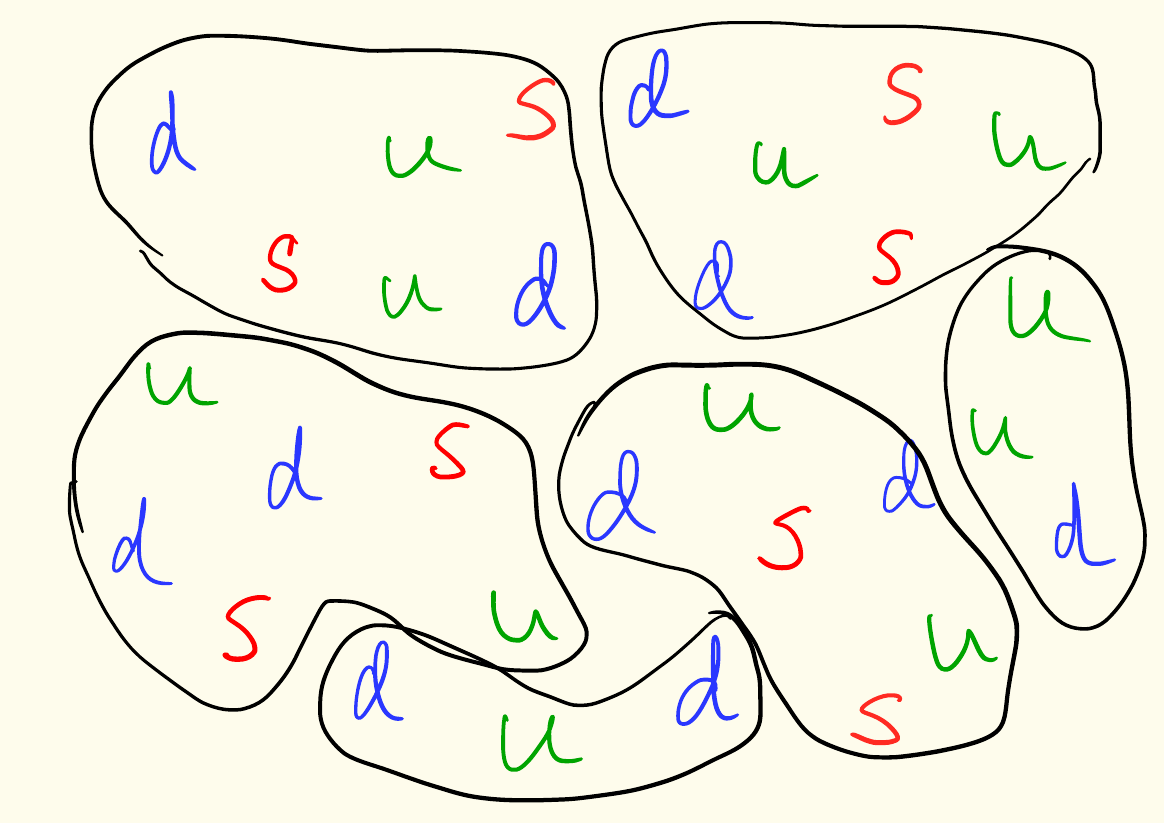}
		\caption{Schematic representation of how left-over $u,d$ quarks make up the baryons if all $s$ quarks are in sexaquark dark matter, for 30\% $s$ quarks;  the spacing is not realistic and the much more abundant background of mesons is not shown.}
	\label{endQGP}
\end{figure} 

\subsection{Comment on arXiv:1803.10242}

Farrar and Zaharijas \cite{fzNuc03} discussed in detail the existing and possible experimental limits on the H-dibaryon mass coming from nuclear stability (sec. IIB).  They concluded that the best limit at that time (2003) was the one which could be inferred from the reported SuperK trigger rate, placing a lower limit on the lifetime of $^{16}O$ to decay to final states which would trigger SuperK, of a few $10^{25}$yr.  (They pointed out analyses the SuperK collaboration could do to potentially improve this limit by several orders of magnitude, but these have not been done.)  They also estimated the $^{16}O$ lifetime as a function of the wavefunction overlap $\mathcal{M}$ between H-dibaryon and two baryons, based on measured weak decay lifetimes, to be (Eq. 36 of ref. [7]):
\be
\tau_{A_{NN} \rightarrow A'_H e \nu} \approx \frac{\kappa_{1440}}{|\mathcal{M}|^2_{\Lambda \Lambda \rightarrow H}}\times 10^5 \, {\rm yr} \, ,
\ee
for $m_H = 1.8$ GeV, where $\kappa_{1440}$ contains the residual color-flavor-spin dependence of the amplitude not accounted for by the simple hadronic-factorization and state-counting model.   

Recently \cite{strumia+18} used the results of \cite{fzNuc03} with the overlap evaluated using equations of \cite{fzNuc03} employing a recent model of the two nucleon wavefunction, and found a lifetime shorter than the bound of  \cite{fzNuc03}, from which they concluded that a stable sexaquark is excluded.  While a 1.8 GeV sexaquark is likely excluded, this does not mean that a stable sexaquark in general is excluded.  There is no constraint on $m_S > 1861$ MeV, for which decay of $^{16}O$ is kinematically forbidden, and somewhat lower $m_S$ may be allowed due to the rough nature of the estimate of \cite{fzNuc03} and the sensitivity of the rate to phase space.   See footnote 3 of the main text for why deuteron stability does not presently provide a stronger constraint. An effort to determine more accurately the allowed mass range for a stable \s\ is underway.

Ref. \cite{strumia+18} also attempts to model the relic density of \s\-DM by assuming thermal freezeout.  For this purpose they take $m_S =1.2$ GeV, in spite of the nuclear-stability exclusion of such low mass.  For this mass, the correct relic abundance is obtained if freezeout occurs at $T=25$ MeV.  They find that the most important reactions to maintain chemical equilibrium are $\Lambda \Lambda \rightarrow S + X$ and others of this type, with cross sections required to be $\mathcal{O}(\rm{GeV}^{-2})$ to maintain equilibrium down to 25 MeV.  However the \s-baryon-baryon overlap required for such a large cross section is excluded, as discussed in \cite{fSexaquark17}.  

In short, neither arXiv:1803.10242's relic abundance calculation nor its exclusion claim based on nuclear stability \cite{fzNuc03} are applicable to the stable sexaquark scenario of \cite{fSexaquark17} and considered here.

\subsection{Direct Detection and Astrophysical and Cosmological limits}

Direct detection searches for hadronically interacting DM (HIDM) must be performed with minimal overburden, to minimize energy-loss from scattering before the detector; limits are presently restricted to $m_{\rm DM} \gtrsim 2-3$ GeV because {\it i)} light DM carries inherently less energy ($KE \sim m_{\rm DM}$) and deposits a smaller fraction of its energy due to mass-mismatch with detector nuclei, {\it ii)} the apportionment of the nuclear recoil energy between heat and production of interstitial defects has not yet been measured \cite{mfVelDep18} and {\it iii)} detectable nuclear recoil events come from the high-velocity tail in the DM heliocentric velocity distribution, which has recently been inferred from GAIA data and found to be significantly lower than in the standard halo model SM\cite{necib+18}.  The best limits on DM in the mass range $\approx 0.6-6 $ GeV are the indirect limits of \cite{nfm18} -- from HST orbital decay and evaporation of liquid cryogens, and thermal conductivity of the Earth -- but these are compatible with DM consisting entirely of \s's, if the mean cross section for DM scattering on a nucleus in the Earth's crust is $\lesssim 10^{-29} \,{\rm cm}^2$;  see \cite{fx18inprep} for a comprehensive discussion.  Note that the above-mentioned direct detection constraints on sexaquark DM only place limits on the product of the fraction of DM in sexaquark form, times the sexaquark-nucleon interaction cross section.  

Limits on quark nuggets and PBH as dark matter are very different in character compared to those for HIDM.  See \cite{a-hKamPBH17} for a review of the constraints on PBH DM; constraints on quark nuggets should be qualitatively similar, although a detailed study is lacking.   An important point to note is that $uds$ ``compact object" DM would have a spectrum of masses, which is harder to constrain than if the mass has a single value.  It is currently not possible to rule out even 100\% of the observable dark matter being PBH (Y.~Ali-Ha{\"i}moud, private commumication), for a general mass distribution.  
Thus until the apportionment of DM between particles and compact objects and the mass distribution of the compact objects is known, it will be difficult to infer limits on the most general $uds$ DM scenario or place limits on the \s-nucleon scattering cross section.

For sufficiently high $\sigma_{DM,p} \gtrsim 10^{-25} {\rm cm}^2$, scattering between DM and gas in galaxies can produce a (thick) DM disk and DM co-rotation, which diminishes the direct DM detection signal significantly for GeV and lighter DM \cite{fSDM17}.  But cross sections sufficient to make an appreciable impact on the local DM distribution appear difficult to reconcile with the limits from \cite{nfm18}, unless only a fraction of DM is particulate, for DM in the sexaquark mass range.   The possible existence of a DM disk is also subject to direct observational constraints, c.f., SM\cite{shavivPaleo16,schutz+17}; and see SM\cite{hooperMcDermott18} for additional discussion.

\subsection{Durability of the \s\ in the hadronic phase}

If breakup processes such as $\pi S \leftrightarrow \Sigma \Lambda$ and $K S \leftrightarrow p \Lambda$ have a typical hadronic rate, they are fast compared to the Hubble expansion rate at the temperatures of interest, $T \sim 150$ MeV, so the \s\ is in chemical equilibrium with baryons and the chemical potentials satisfy $\mu_S = 2 \mu_b$.  In this case, an initial SDM excess comparable to the baryon excess would quickly disappear, because $\mu_S = 2 \mu_b$ implies that in chemical equilibrium $\lesssim 10^{-7}$ of the baryon number is carried by \s\ 's at $T \sim 150$ MeV, assuming $m_S \approx 2 m_p$.  

The breakup rate can be calculated using lowest order meson-baryon effective field theory, extended to include an \s\ and have an $SBB'$ vertex.  Projecting the \s\ wavefunction in terms of quarks onto octet baryons, gives the color-flavor-spin wavefunction-overlap  and hence the relative couplings of the \s\ to various $BB'$ flavors\cite{wintergerst15}:
\begin{align}
\label{SBB'}
<\, S \, | \,\Lambda \Lambda> & = \,<\, S \, | \,\Sigma^0 \Sigma^0> \, = - <\, S \, | \,\Sigma^+ \Sigma^->  \\
					& =  - <\, S \, | \, n \, \Xi^0  > \, = \, <\, S \, | \, p \, \Xi^-  > = \frac{\tilde{g}}{\sqrt{40}}, \nonumber
\end{align}
where $\tilde{g}$ is the effective field theory coupling reflecting the spatial wavefunction overlap between \s\ and baryons.  The $\frac{1}{\sqrt{40}}$ reflects the fact that products of color singlets make up only 20\% of the \s\ wavefunction.  

The breakup processes with the highest rates are $\pi^\pm S \rightarrow \Sigma^\pm \Lambda$ and $K^+ S \rightarrow p \Lambda$ with amplitudes 
\be
\label{ampLamSig}
\mathcal{M}_{\pi^\pm S \rightarrow \Sigma^\pm \Lambda} \approx \frac{f \tilde{g}}{\sqrt{120}} (1-\alpha) m_\Lambda m_\Sigma v_{\rm rel} \left(\frac{1}{m_\Lambda^2} - \frac{1}{m_\Sigma^2} \right) ;
\ee
\be
\mathcal{M}_{K^+ S \rightarrow p \Lambda} \approx \frac{f \tilde{g}}{\sqrt{120}} m_\Lambda m_p v_{\rm rel} \left( - \frac{(1+ 2 \alpha)}{m_\Lambda^2}  -  \frac{(4 \alpha - 1)}{m_\Xi^2} \right)  .\nonumber
\ee
Here $f = 0.952$ and $\alpha = 0.365$ are taken from \cite{stoksMBcouplings97} where these parameters characterizing the meson-baryon couplings are fit to data and $v_{\rm rel}$ is the relative velocity in the final state.   The $v_{\rm rel}$ factor arises because the baryons must have $L=1$ in order to satisfy parity and angular momentum conservation and Fermi statistics, given that the $\pi/K$ is a pseudoscalar, the \s\ is an even parity, spin-0 particle, and the intrinsic parity of a pair of baryons is +1.   Note that $\pi^\pm S \rightarrow \Sigma^\pm \Lambda$ vanishes in the flavor SU(3) limit due to the opposite sign of $<\, S \, | \,\Lambda \Lambda>$ relative to $<\, S \, | \,\Sigma^+ \Sigma^->$ in Eq. (\ref{SBB'}).   

The effective coupling $\tilde{g}$ is simply the transition amplitude between 6 quarks in the \s\ and in two baryons, since the flavor-spin factors have already been included; this is just the overlap of the spatial wavefunctions.  Ref. \cite{fzNuc03} evaluated the geometrical overlap for several models of the baryon wavefunctions, not including a suppression for tunneling through any potential barrier.  The results are shown in Fig \ref{overlap}.   The parameter $f = \frac{r_B}{r_S}$ with $r_B = 0.9$ fm and (see main text) $r_S = \lambda_S + (0-0.5) \lambda_{M1}$ .  Since the $f_0$ is either an $\ell = 1$ $q\bar{q}$ or a tetraquark or a bound-state of two pions it couples weakly to the \s\ so the \s's size is governed by its coupling to the flavor-singlet combination of $\omega$ and $\phi$ mesons with mass $\approx 1$ GeV.   That leads to $r_S  = 0.1-0.2$ fm and $f = 4.5-9$.

\begin{figure}[b!]
	\vspace{-0.1in}
	\label{overlap}
	\centering
	\includegraphics[trim = 0.2in 0.0in 0.2in 0.4in, clip, width=0.48\textwidth]{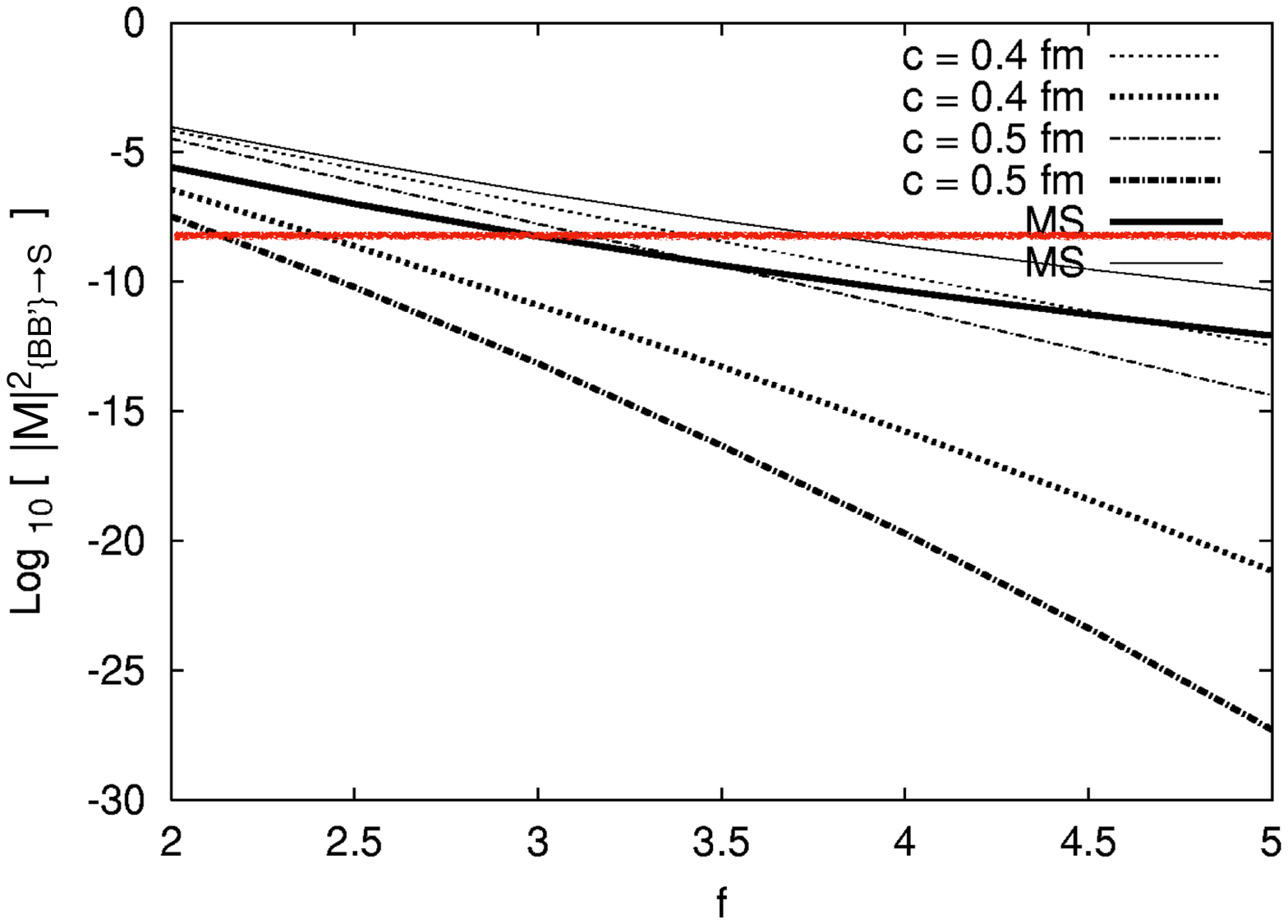}
	\vspace{-0.45in}
	\caption{Wavefunction overlap $\tilde{g}^2$ between \s\ and two baryons in a nucleus, as a function of the ratio $f=\frac{r_B}{r_S}$, for three different nuclear wavefunctions.   The heavy lines are for the standard Isgur-Karl value of the size parameter $\alpha_B = 0.406 \,{\rm fm}^{-1}$. Results for a comparison value $\alpha_B = 0.221 \,{\rm fm}^{-1}$ were given in \cite{fzNuc03}, for completeness.  The red fuzzy line is the approximate upper limit from doubly-strange hypernuclei.  The effective coupling $\tilde{g}^2$ is this wavefunction-overlap-squared times any tunneling suppression.  Adapted from \cite{fzNuc03}.}\label{overlap}
	\vspace{-0.2in}
  \end{figure}

Performing the thermal average following \cite{cannoni16}, gives the \s\ breakup rates for these two channels.  $\Gamma(K^+ S \rightarrow p \Lambda) = n_{K^+}(T) <\sigma_{K^+ S \rightarrow p \Lambda} v>$ is about two orders of magnitude larger than $\Gamma(\pi^\pm S \rightarrow \Sigma^\pm \Lambda)$, the suppression of the latter resulting from the cancelation between the contributions of virtual $\Lambda$ and $\Sigma$ in (\ref{ampLamSig}).   

The Hubble expansion rate is greater than $\Gamma(K^+ S \rightarrow p \Lambda) = n_{K^+}(150 {\rm MeV}) <\sigma v>$, for $\tilde{g}^2 <  4 \times 10^{-12} $.   As can be seen from Fig. \ref{overlap} this is comfortably in the expected range, even without considering a tunneling suppression, since $f = 4.5 - 9$.

\end{document}